\documentclass[aps,pre,showpacs,floatfix,superscriptaddress,twocolumn]{revtex4-1}
\usepackage{graphicx}
\usepackage{amsmath,amssymb,amsthm}
\usepackage{mathptmx}
\usepackage{times}
\usepackage{bm}
\usepackage{color}

\newcommand{\be}{\begin{eqnarray}}
\newcommand{\ee}{\end{eqnarray}}
\newcommand{\wdh}[1]{\hat{#1}}
\def\beq{\begin{equation}}
\def\eeq{\end{equation}}

\begin{document}

\title{Quantum quenches in disordered systems: \\
Approach to thermal equilibrium without a typical relaxation time}

\author{Ehsan Khatami}
\affiliation{Department of Physics, Georgetown University, Washington, DC 20057, USA}
\author{Marcos Rigol}
\affiliation{Department of Physics, Georgetown University, Washington, DC 20057, USA}
\author{Armando Rela\~no}
\affiliation{Departamento de F\'{\i}sica Aplicada I, 
Universidad Complutense de Madrid, Av. Complutense s/n, 28040 Madrid, Spain}
\affiliation{Instituto de Estructura de la Materia, IEM-CSIC, Serrano 123, 28006 Madrid, Spain}
\author{Antonio M. Garc\'{\i}a-Garc\'{\i}a}
\affiliation{CFIF, Instituto Superior T\'ecnico, 
Universidade T\'ecnica de Lisboa, Av. Rovisco Pais, 1049-001 Lisboa, Portugal}
\affiliation{TCM Group, Cavendish Laboratory, University of Cambridge, JJ Thomson Avenue,
Cambridge, CB3 0HE, UK}
\begin{abstract}
We study spectral properties and the dynamics after a quench of one-dimensional spinless 
fermions with short-range interactions and long-range random hopping. We show that a sufficiently fast decay 
of the hopping term promotes localization effects at finite temperature, which prevents thermalization even if 
the classical motion is chaotic. For slower decays, we find that thermalization does occur. However, 
within this model, the latter regime falls in an unexpected universality class, namely, observables exhibit 
a power-law (as opposed to an exponential) approach to their thermal expectation values.
\end{abstract}

\maketitle

The study of the dynamics and conditions for thermalization 
in isolated many-body systems has a long history in theoretical physics \cite{boltzman}. In classical physics, 
the requirements for thermalization are well understood. Boltzmann's ergodic hypothesis holds only if the motion 
of the individual particles is fully chaotic. The situation in quantum systems is less clear. Quantum time 
evolution is described by a linear differential equation so that dynamical chaos does not occur. The development 
of the theory of quantum chaos in the 1980's and 1990's brought a new language and techniques to tackle this problem 
\cite{stock}. With respect to thermalization, this effort crystallized in two main results: the so-called Berry's 
hypothesis \cite{berry}, which states that eigenstates of a classically chaotic system can be written as a finite 
sum of plane waves with random coefficients, and the eigenstate thermalization hypothesis (ETH) proposed by Deutsch 
\cite{deutsch91} and Srednicki \cite{srednicki94}, which states that observables in individual eigenstates 
of a generic many-body system already exhibit thermal behavior. 
 
Until a few years ago, technical difficulties prevented a systematic study of the proposals above. 
However, recent advances in cold gases systems, together with the enhancement of computer power and the development 
of novel computational methods, are making possible a more quantitative comparison. This has dramatically boosted 
the theoretical and experimental interest in non-equilibrium dynamics in general and thermalization in 
particular \cite{cazalilla}. For instance, in Ref.~\cite{kinoshita06}, it was shown experimentally that, 
after a quench, the momentum distribution of a gas of bosons trapped in a quasi-one-dimensional (1D) geometry did not 
relax to the thermal prediction. Although this is expected in an integrable system  \cite{rigol07STAT}, it was 
surprising that such an effect could be seen experimentally. The ETH, on the other hand, has been confirmed 
numerically for nonintegrable systems \cite{rigol08STAT}, and has been shown to break down when approaching 
integrable points \cite{rigol09STAT,santos,Neuenhahn10}.

A better understanding of the conditions for thermalization would not only put the foundation of quantum statistical 
mechanics on a more solid ground but also have a strong impact in different fields, from cold gases to 
cosmology, where non-equilibrium dynamics play a key role. Here, we provide further insights on this problem. We show 
that many-body localization effects \cite{huse} invalidate the expectation that classical chaos leads 
to thermalization of the quantum counterpart. (For recent connections between the effect of localization in 
Fock space and thermalization in spin systems, see Ref.~\cite{fazio}.) We also put forward a route toward 
thermalization in quantum systems characterized by power-law relaxation dynamics. We support these results by 
numerical calculations in the following 1D spinless fermions system, 
with long-range hopping and short-range interactions,
\begin{equation}
\wdh{H} = \sum_{ij} J_{ij} \left( \wdh{f}^{\dagger}_i \wdh{f}_j + \text{H.c.} \right) + 
V\sum_i \left( \wdh{n}_i - \frac{1}{2} \right) \left( \wdh{n}_{i+1} - \frac{1}{2} \right),
\label{eq:model}
\end{equation}
where $\wdh{f}^{\dagger}_j$ creates a fermion in the site $j$, and
$\wdh{n}_j$ is the number operator in the site $j$. The hopping term is built by means of a 
Gaussian random distribution, with zero mean $\langle J_{ij} \rangle=0$, and 
\begin{equation}
\left< \left( J_{ij} \right)^2 \right> = \left[1+\left( \frac{|i-j|}{\beta} \right)^{2 \alpha}\right]^{-1}.
\end{equation}

In the limit $V=0$, the properties of (\ref{eq:model}) depend on $\alpha$ but not on $\beta>0$ \cite{mirlin}. 
For $\alpha<1$, eigenstates are delocalized and the spectral correlations are described by Wigner-Dyson (WD) statistics. 
For $\alpha>1$, eigenstates are localized and spectral correlations are described by Poisson statistics. 
For $\alpha=1$, eigenstates are multifractal  
and spectral correlations are intermediate between WD and Poisson \cite{nishigaki,sko}. 
We then fix $\beta=0.1$ and $V=1$. (The latter sets the unit of energy
throughout this Rapid Communication, and $k_B=1$.) 
These values were chosen to minimize finite-size effects ($\beta \ll 1$) and at the same time to avoid the trivial 
insulator limit that occurs if the potential energy is much larger than the kinetic one \cite{santos}. Likewise, 
if the interaction is too weak, thermalization may not occur
\cite{rigol09STAT}. Finally, we note that these types of long-range Hamiltonians
have been used to model systems with strong dipolar interactions \cite{levitov}.

We first use a finite-size scaling analysis to investigate the effect of the many-body interactions on the 
spectrum \cite{sko}. This is a powerful tool to study many-body localization in the presence of 
interactions \cite{huse,santos}. We compute the eigenvalues of the Hamiltonian \eqref{eq:model} for different sizes 
and values of $\alpha$ utilizing standard diagonalization techniques. In all cases, the filling factor $p/L=1/3$, 
where $L$ is the system size and $p$ is the number of particles. The spectrum thus obtained is appropriately 
unfolded, i.e., it is rescaled so that the spectral density on a spectral window 
comprising several level spacings is unity. The number of disorder realizations considered 
is such that statistical fluctuations are negligible. 
As a scaling variable we choose the function $\eta$ \cite{sko} related to the variance 
${\rm var} =\langle s^2\rangle-\langle s \rangle^2$ of the level spacing distribution $P(s)$. 
$P(s)$ is the probability of finding two neighboring eigenvalues at 
a distance $s = (\epsilon_{i+1} - \epsilon_{i})/\Delta$, where $\Delta$ is the local 
mean level spacing, and
\begin{equation}
\eta = [{\rm var} - {\rm var_{WD}}\,] / [\,{\rm var_{P}}-{\rm var_{WD}}] \;.
\label{eta}
\end{equation}
${\rm var_{WD}} = 0.286$ $({\rm var_P} = 1)$ is the value of the variance for a disordered metal 
(insulator) in the thermodynamic limit 
and $\langle s^n \rangle = \int s^n P(s)$ \cite{C99}.
An increase (decrease) of $\eta$ with $L$ signals that the system will be an insulator (metal) 
in the thermodynamic limit.

\begin{figure}[!t]
\includegraphics[width=0.85\columnwidth,clip]{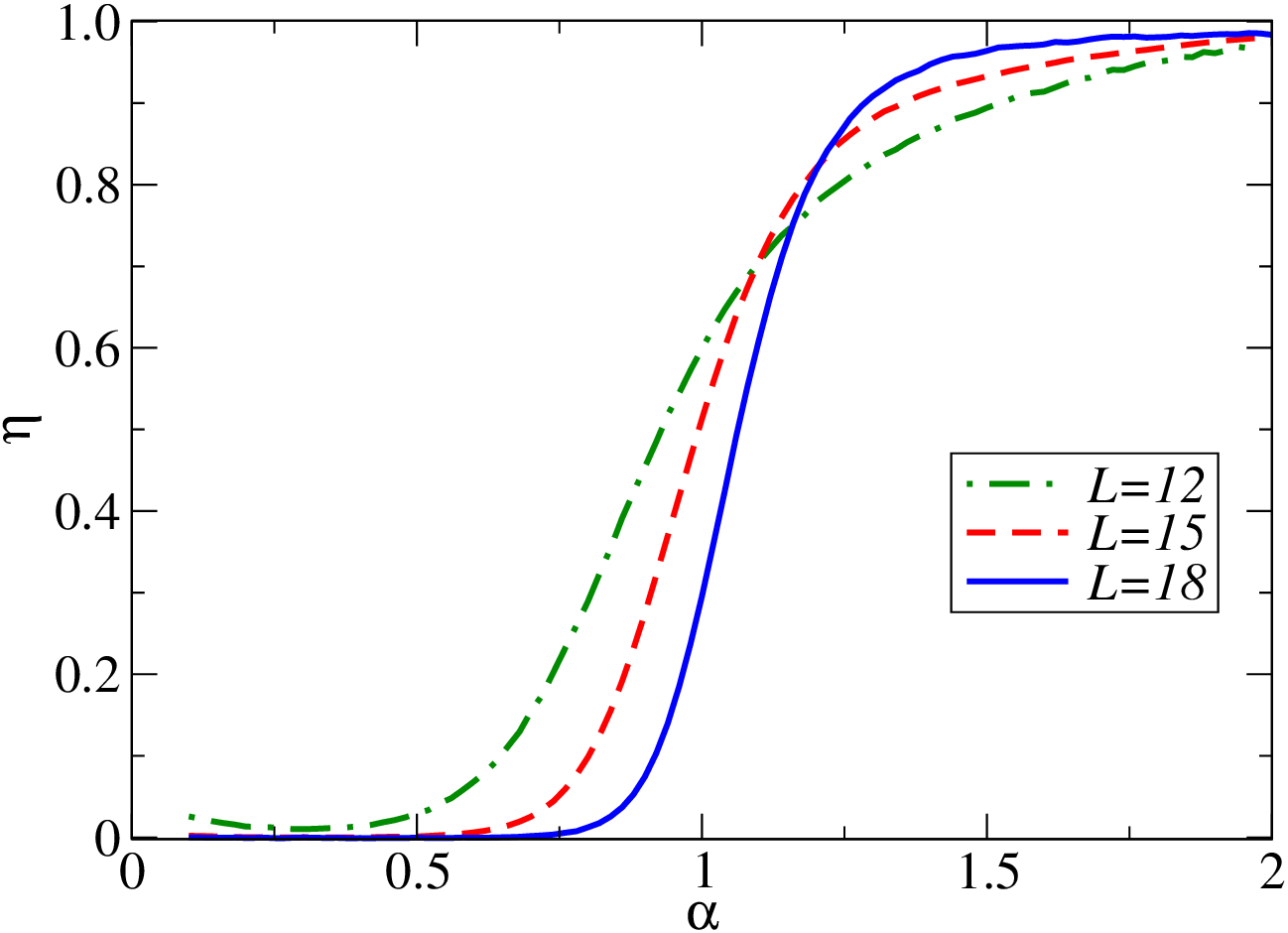}
\vspace{-0.3cm}
\caption{(Color online) Scaling variable $\eta$ [see Eq.~\eqref{eta}] as a function of $\alpha$ for 
different system sizes but the same filling factor, $1/3$. For $\alpha \lesssim 1$ ($\alpha \gtrsim 1.2$), 
$\eta$ decreases (increases) with the system size. This is a signature of a metallic (insulating) phase.
The number of realizations is $10000$, $10000$, and $400$ for $L=12$, $15$, and $18$, respectively.}  
\label{fig1} 
\end{figure}

Figure \ref{fig1} depicts results for $\eta$ vs $\alpha$ for different sizes. It is apparent that for 
$\alpha \gtrsim 1.2$ 
the parameter $\eta$ increases with system size, as is expected of an insulator. Hence, localization
takes place in the interacting system, in contrast to the classical counterpart,
which is chaotic for any $\alpha$. For $\alpha \lesssim 1$, on the other hand, 
$\eta$ behaves as expected of a metal. To be certain whether the system is metallic 
for $\alpha \approx 1$, much larger systems, currently not accessible numerically, need to be studied.

We now investigate the thermalization properties of the Hamiltonian (\ref{eq:model}). We aim to 
(i) identify a region of $\alpha$'s for which the system does not thermalize even though the classical 
counterpart does, (ii) see how that region relates to the localization regime found by the spectral analysis, 
(iii) investigate the microscopic origin of the lack of thermalization, and (iv) study the approach to 
equilibrium in the region of $\alpha$'s for which thermalization eventually occurs. 

In order to proceed, we first note that time scales and 
finite-size effects relevant to the study of quantum thermalization may depend on the observable and particle 
statistics \cite{rigol09STAT,Neuenhahn10}. However, for few-body observables, it is generically expected that thermalization 
occurs away from integrability. Here, we report results for two of those observables: the momentum distribution 
function [$\hat{n}(k)$] and the density-density structure factor [$\hat{N}(k)$],
\begin{equation}
\hat{n}(k)=\dfrac{1}{L}\sum_{l,m} e^{ik(l-m)} 
\hat{f}^{\dagger}_l\hat{f}^{}_m,\ \ \hat{N}(k)=\dfrac{1}{L}\sum_{l,m} e^{ik(l-m)} 
\hat{n}^{}_l\hat{n}^{}_m,
\end{equation}
which are the Fourier transforms of the one-particle and density-density correlation matrices \cite{note1}, 
respectively. Both observables can be measured in ultracold gases experiments.

In all the cases shown below, we start from an eigenstate of the Hamiltonian (\ref{eq:model}) in a certain
realization of the disorder. Then, we change to another disorder realization for
the same $\alpha$, and study the time evolution of the initial state with this
final Hamiltonian,
$\hat{H}_\textrm{fin}$. This procedure is usually known as a quench. The initial
state [$|\Psi(0)\rangle$]
is selected such that the time evolving system has an energy 
$E=\langle\Psi(0)|\hat{H}_\textrm{fin}|\Psi(0)\rangle$, which, 
for every disorder realization, is the same as the one of a thermal state with temperature $T=10.0$ 
($E=\text{Tr}\{e^{-\hat{H}_\textrm{fin}/T}\hat{H}_\textrm{fin}\}/\text{Tr}\{e^{-\hat{H}_\textrm{fin}/T}\}$). 
This yields energies that fall close to the center of the spectrum of the final Hamiltonian. In what follows, 
$O_{ij}$ are the matrix elements of a given observable in the eigenstates of the final 
Hamiltonian, $O_{ij}=\langle \psi_i| \hat{O}|\psi_j \rangle$, and $C_j$ is the overlap between the initial 
state and $|\psi_j \rangle$, $C_j = \langle \psi_j | \Psi(0) \rangle$. 

In order to determine whether thermalization occurs following the quench, one needs to find a meaningful 
quantity to compare with the microcanonical (thermal) average, 
$O_\textrm{micro} = \frac{1}{{\cal N}_{\Delta E}}\sum_j O_{jj}$, where ${\cal N}_{\Delta E}$ is the number
of states in the microcanonical window (centered around $E$, and with $\Delta E$ selected such that 
the average is robust against small changes of $\Delta E$). If relaxation takes place for the observables of interest, 
and the spectrum is nondegenerate, the infinite time average (also known as the diagonal ensemble prediction) 
$O_\textrm{diag} = \sum_j |C_j|^2 O_{jj}$ is the right choice \cite{rigol08STAT}. We first compute the  
normalized difference between these two ensembles,
\be 
\label{terma}
\Delta O =\frac{\sum_k |O_\textrm{diag}(k) - O_\textrm{micro}(k)|}{\sum_k O_\textrm{diag}(k)},
\ee
and then average it over different disorder realizations to obtain $\langle\Delta O\rangle_\textrm{dis}$. Note
that here, and in what follows, by ``$O$'' we mean ``$n$'' or ``$N$'' for the momentum distribution and structure 
factor, respectively.
\begin{figure}[!t]
\includegraphics[width=0.85\columnwidth,clip]{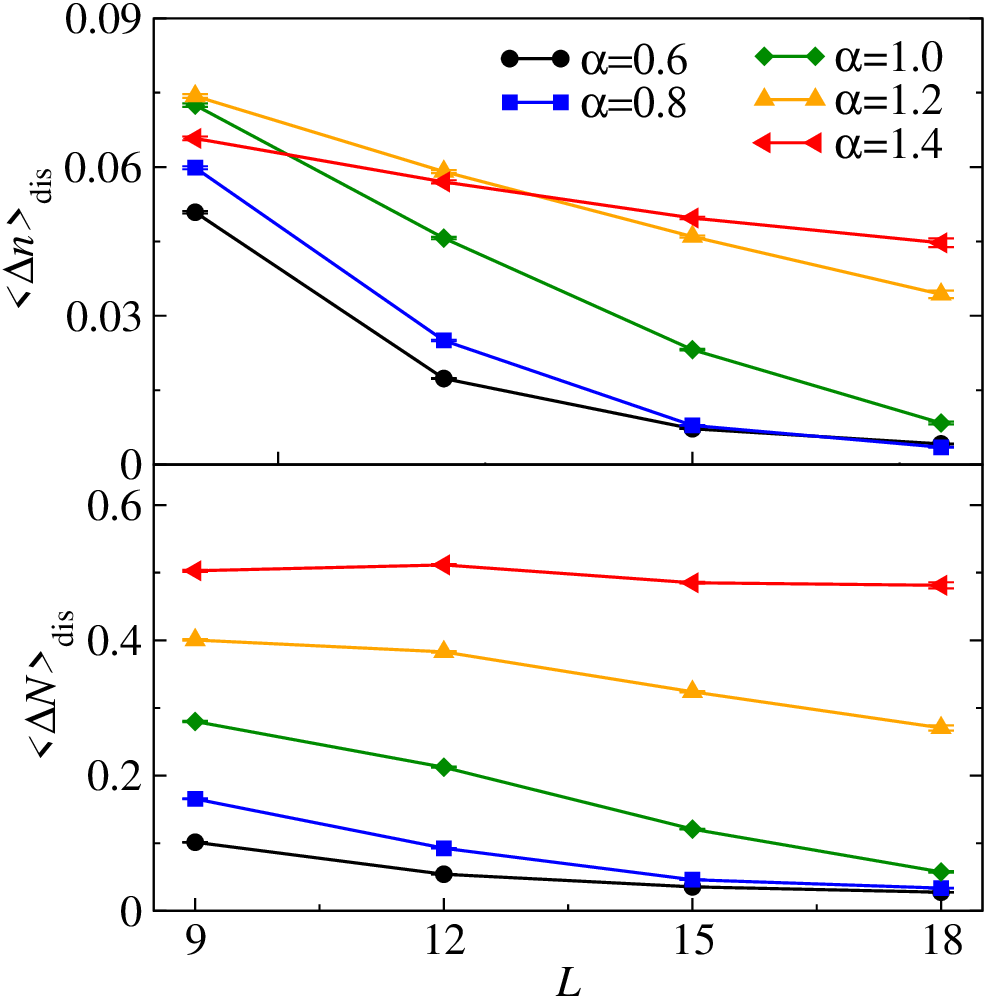}
\vspace{-0.3cm}
\caption{(Color online) $\langle\Delta n\rangle_\textrm{dis}$ and $\langle\Delta N\rangle_\textrm{dis}$ 
[see Eq.~\eqref{terma}], as a function of the system size, for $\alpha=0.6$, $0.8$, $1.0$, $1.2$, and $1.4$. 
Points correspond to $(L,p) = (9,3), (12,4)$, $(15,5)$, and $(18,6)$. The disorder 
average is performed over at least $8,500$ different realizations for $L\le15$, and $1020$ for $L=18$. 
The classical dynamics is chaotic but there is no thermalization for large $\alpha$ due to localization effects.}  
\label{fig2} 
\end{figure}

\begin{figure}[!b]
\includegraphics[width=0.95\columnwidth,clip]{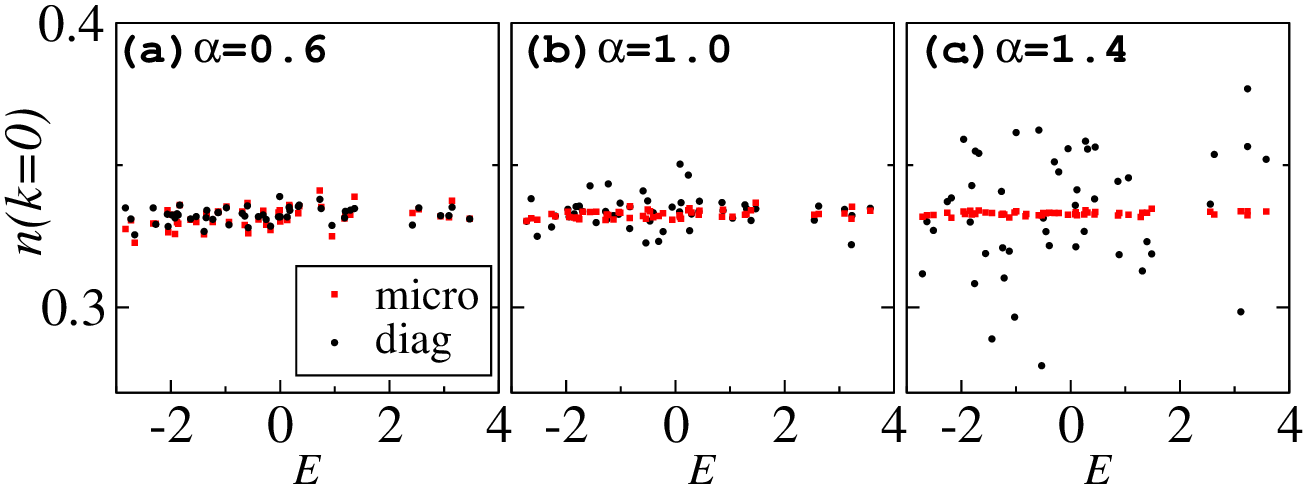} 
\vspace{-0.3cm}
\caption{(Color online) Microcanonical and diagonal results for $n(k=0)$ in $51$ different 
quenches. The systems have $18$ sites and six particles, with $\alpha=0.6$, $1.0$, and 
$1.4$.}
\label{fig:extra} 
\end{figure}

In Fig.~\ref{fig2}, we depict $\langle\Delta n\rangle_\textrm{dis}$ and $\langle\Delta N\rangle_\textrm{dis}$ 
for different values of $\alpha$ vs system size. Thermalization occurs if $\langle\Delta O\rangle_\textrm{dis}$ 
vanishes in the thermodynamic limit. A nonzero value of $\langle\Delta O\rangle_\textrm{dis}$ in this limit indicates 
that the observable $O$ relaxes to a nonthermal expectation value. For $\alpha > 1$, a weak size dependence is 
observed for the largest system sizes we can study, with $\langle\Delta O\rangle_\textrm{dis}$ likely saturating
to non-zero values for $\alpha \gtrsim 1.2$. Therefore, thermalization is not expected to occur in this regime.
Interestingly, as the value of $\alpha$ decreases below $\alpha \sim 1$, the normalized differences exhibit a 
fast decrease for the smallest systems shown. They become much smaller than those for $\alpha \gtrsim 1.2$ for the largest 
system sizes accessible here, for which $\langle\Delta O\rangle_\textrm{dis}$ is very close to zero within our error bars
and still decreasing with increasing system size. These results suggest that thermalization occurs in this regime.

In order to further support the conclusions of the finite-size scaling analysis we look at the actual diagonal and 
microcanonical expectation values of observables 
for several quenches. Results for $n(k=0)$ are shown in Fig.~\ref{fig:extra} as a function of the energy. 
In all regimes, the microcanonical results can be seen to be almost independent of the energy, while the diagonal ensemble
results exhibit fluctuations that increase as $\alpha$ increases. Hence, increasing $\alpha$ increases the difference 
between the infinite time average and the microcanonical results, as well as increases the dependence of the  
infinite time average on the initial state selected.

A natural question that follows is whether the absence of thermalization, as well as the sensitivity to the initial
state selected, for large $\alpha$, is related to the breakdown of ETH (as seen in clean systems approaching an integrable
point \cite{rigol09STAT,santos}) or it is rather related to some atypical properties of the overlaps $C_j$. 
To answer that question, we compute the normalized difference between the observable in each eigenstate and 
in the microcanonical ensemble,
\be 
\label{eq:ext}
\Delta O_{ii} = \frac{\sum_k |O_{ii}(k) - O_\textrm{micro}(k)|}
{\sum_k O_\textrm{micro}(k)}.
\ee
This allows us to determine, for each disorder realization, 
the maximal difference within the microcanonical window $\Delta O_{ii}^\textrm{max}=\text{Max}[\Delta O_{ii}]_{\Delta E}$ 
as well as the average $\Delta O_{ii}^\textrm{avg}=\frac{1}{\mathcal{N}_{\Delta E}}\sum_i \Delta O_{ii}$. We then
average both quantities over different disorder realizations to obtain 
$\langle\Delta O_{ii}^\textrm{max}\rangle_\textrm{dis}$ and 
$\langle\Delta O_{ii}^\textrm{avg}\rangle_\textrm{dis}$.

\begin{figure}
\includegraphics[width=0.85\columnwidth,clip]{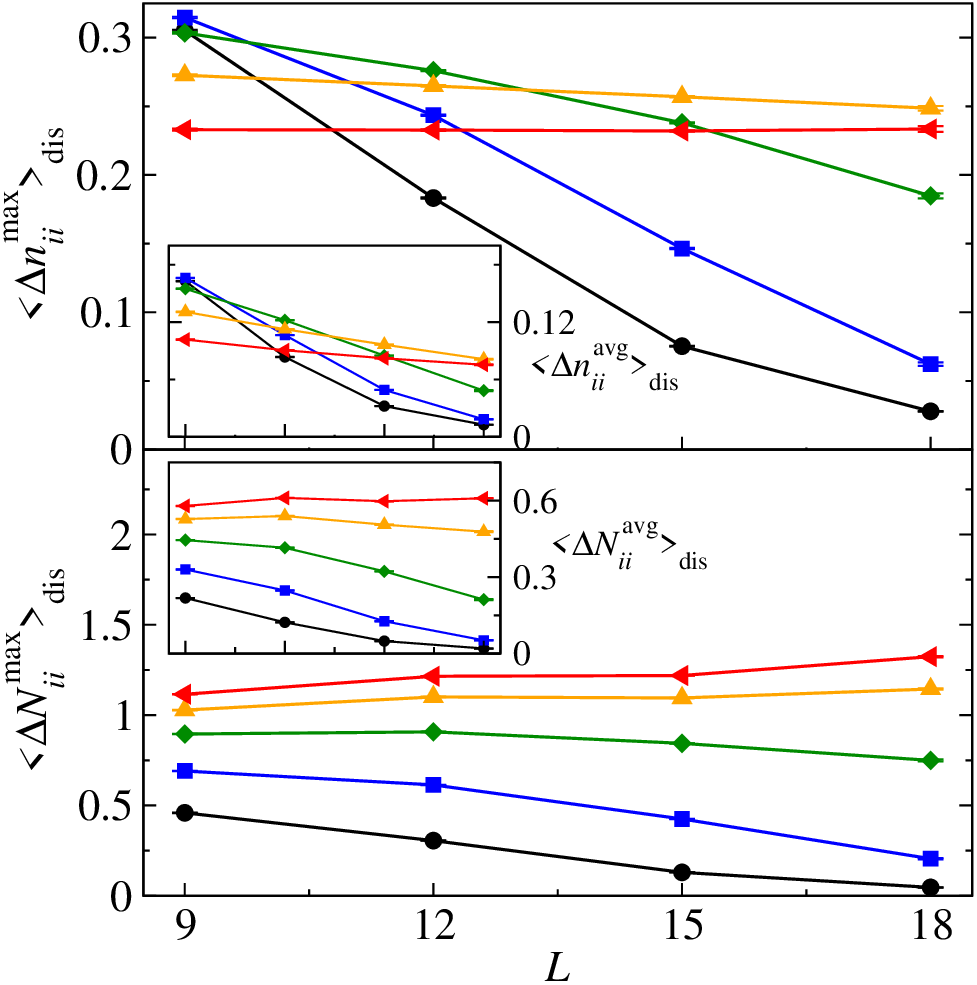}
\vspace{-0.3cm}
\caption{(Color online) $\langle\Delta n_{ii}^\textrm{max}\rangle_\textrm{dis}$ and
$\langle\Delta N_{ii}^\textrm{max}\rangle_\textrm{dis}$ (main panels), and 
$\langle\Delta n_{ii}^\textrm{avg}\rangle_\textrm{dis}$ and
$\langle\Delta N_{ii}^\textrm{avg}\rangle_\textrm{dis}$ (insets) vs system size. 
Lines are the same as in Fig.~\ref{fig2}. As a consequence of localization effects, 
ETH does not hold for large values of $\alpha$.}
\label{fig3} 
\end{figure}

In the main panels in Fig.~\ref{fig3}, we depict $\langle\Delta n_{ii}^\textrm{max}\rangle_\textrm{dis}$ and
$\langle\Delta N_{ii}^\textrm{max}\rangle_\textrm{dis}$ vs $L$ for different values of $\alpha$.
ETH holds when $\langle\Delta O_{ii}^\textrm{max}\rangle_\textrm{dis} \to 0$ for $L\rightarrow\infty$. 
In the range of sizes that we can study, this behavior is apparent for $\alpha \lesssim 1$. For $\alpha \gtrsim 1.2$, 
we find clear indications that ETH fails, which can be understood as a result of localization induced 
by disorder \cite{huse}. A very similar behavior is observed in the insets of Fig.~\ref{fig3}, which show 
$\langle\Delta n_{ii}^\textrm{avg}\rangle_\textrm{dis}$ and
$\langle\Delta N_{ii}^\textrm{avg}\rangle_\textrm{dis}$. In the region $\alpha \approx 1$, on the other hand, 
our results are not conclusive. This is 
an interesting problem for future works as, in the noninteracting limit, $\alpha =1 $ corresponds to a 
metal-insulator transition characterized by multifractal eigenstates. We speculate that fluctuations 
at all scales associated with multifractality may lead to interesting behavior in the many-body 
properties of the system. 

The robustness of the results for $\langle\Delta O_{ii}^\textrm{max}\rangle_\textrm{dis}$ and 
$\langle\Delta O_{ii}^\textrm{avg}\rangle_\textrm{dis}$, as well as their clear correlation with 
the thermalization indicators in Fig.~\ref{fig2}, allow us to conclude that: (i) the lack (occurrence) of 
thermalization is directly related to the failure (validity) of ETH, and (ii) that ETH holds and thermalization 
occurs only for values of $\alpha \lesssim 1$. For $\alpha$ greater than, and not too close to, 
one, the quantum system will not thermalize even though the dynamic of the classical counterpart 
is chaotic.

\begin{figure}[!t]
\includegraphics[width=0.9\columnwidth,clip]{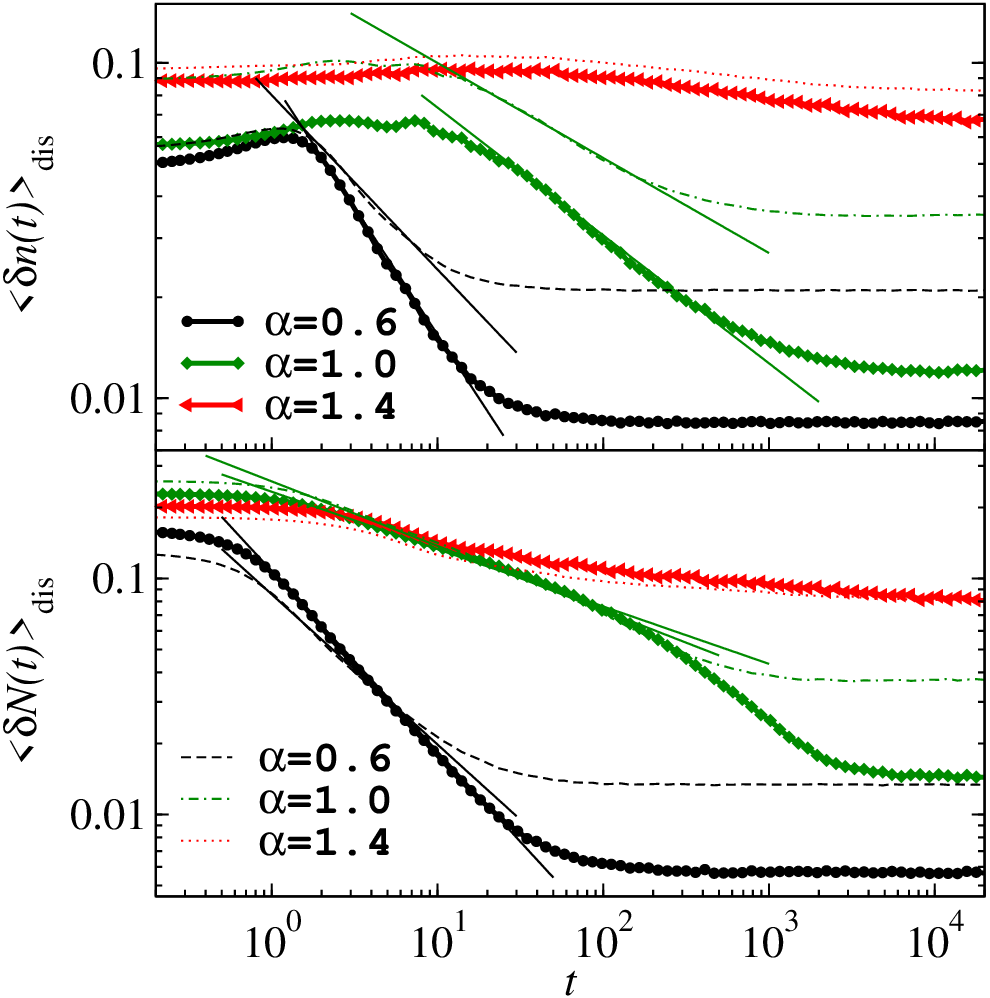} 
\vspace{-0.3cm}
\caption{(Color online) Time evolution of $\langle\delta n(t)\rangle_\textrm{dis}$ and 
$\langle\delta N(t)\rangle_\textrm{dis}$ [see Eq.~\eqref{Eq:errorN}] for different 
$\alpha$'s. Thick lines are for $18$ sites and six particles. Thin solid lines are 
power-law fits to the data. Other thin lines are the corresponding results for $15$ sites 
and five particles. An average over $8500$ ($1020$) realizations has been carried out
for the 15-site (18-site) system.}  
\label{fig4} 
\end{figure}

A fundamental problem that has not been addressed in previous computational studies--due to large 
fluctuations that occur during the time evolution in exact diagonalization studies (because of finite-size 
effects) and to the limited times accessible in time-dependent 
density-matrix renormalization group (t-DMRG) approaches \cite{dmrg}--is that of how observables approach 
their thermal values during the relaxation dynamics. The naive expectation is that the approach should be 
exponential as in classical systems, where a few collisions per particle suffice for the system to relax 
to thermal equilibrium. However, to the best of our knowledge, such a behavior has yet to be seen in the 
experiments or numerical 
simulations of isolated systems in the quantum regime. Disordered systems provide a unique opportunity to 
address this problem as the average over disorder realizations reduces dramatically fluctuations due to 
finite-size effects. In what follows, we compute the time evolution of the difference 
\begin{equation}
 \delta O(t)=\dfrac{\sum_k|O(k,t)-O_{\textrm{diag}}(k)|}{\sum_k O_{\textrm{diag}}(k)},
\label{Eq:errorN}
\end{equation}
and then average it over different disorder realizations to obtain $\langle\delta O(t)\rangle_\textrm{dis}$ 
\cite{rigol09STAT}. In Fig.~\ref{fig4}, we show results for $\langle\delta n(t)\rangle_\textrm{dis}$ and 
$\langle\delta N(t)\rangle_\textrm{dis}$ as a function of time, $t$, for three different values of $\alpha$ 
and the two largest system sizes that we can study. For $\alpha>1$, an extremely slow relaxation dynamics
can be observed, and the system may never reach the infinite time-average prediction in any experimentally 
relevant time scale. For $\alpha\lesssim1$, the relaxation dynamics seen in those plots is quite unexpected. 
We find that $\langle\delta n(t)\rangle_\textrm{dis}$ and $\langle\delta N(t)\rangle_\textrm{dis}$ exhibit 
a clear power-law decay ($\propto t^{-\gamma}$). The region over which the power-law decay is observed extends 
over a decade and increases with increasing system size. As the value of $\alpha$ decreases (and localization 
effects decrease) the exponent $\gamma$ of the power law increases. However, no typical time-scale emerges 
during relaxation \cite{note2}. This indicates an unexpected route to thermal equilibrium in many-body quantum 
systems characterized by a power law rather than an exponential decay.

After these results, it is natural to speculate whether such power-law behavior also occurs in clean systems. 
Theoretically, it is well known that in the semiclassical limit classical
cantori \cite{cantori}, remnants of the Kolmogorov-Arnold-Moser (KAM) 
tori induce slow diffusion in phase space and power-law localization of eigenstates in the one-body problem 
\cite{antwan}. Therefore, it is plausible that in certain region of parameters the approach to equilibrium in 
systems controlled by cantori is also power-law-like.
Interestingly, indications of power-law relaxation have already been
seen in classical systems \cite{alder} and, experimentally,  in a strongly
correlated one-dimensional Bose gas \cite{uli}.

In conclusion, we have studied an interacting many-body disordered system that
exhibits a transition between metallic and insulating behavior. Remarkably, we have identified 
a region of parameters ($\alpha \gtrsim 1.2$) in which, 
due to localization effects, ETH fails and thermalization does not take place even if the system is 
nonintegrable \cite{fazio}. For $\alpha\lesssim1$, ETH is valid and thermalization occurs. 
Furthermore, in the latter regime, we have found a route toward thermal equilibrium characterized 
by a power-law approach to the thermal expectation values and, hence, by the lack of a well-defined 
equilibration time. The relevance of this scenario to experiments with ultracold gases, as well as
clean strongly correlated systems, is a topic that requires future exploration.

\vspace{0.1in}
This research was supported by NSF under Grant No.~OCI-0904597 (E.K. and M.R.) 
and by the U.S. Office of Naval Research (M.R.). A.M.G. acknowledges support from  Galileo Galilei Institute, 
FCT (PTDC/FIS/111348/2009), Marie Curie Action (PIRG07-GA-2010-26817), and EPSRC (EP/I004637/1). 
A.R. acknowledges support from the Spanish Government Grants No.~FIS2009-11621-C02-01 and No. FIS2009-07277.

\vspace{-0.1in}
% References %%%%%%%%%%%%%%%%%%%%%%%%%%%%%%%%%%%%%%%%%%%%%%%%%%%%%%%%%%%%%%%%%%%%%


\begin{thebibliography}{99}
\expandafter\ifx\csname natexlab\endcsname\relax\def\natexlab#1{#1}\fi
\expandafter\ifx\csname bibnamefont\endcsname\relax
  \def\bibnamefont#1{#1}\fi
\expandafter\ifx\csname bibfnamefont\endcsname\relax
  \def\bibfnamefont#1{#1}\fi
\expandafter\ifx\csname citenamefont\endcsname\relax
  \def\citenamefont#1{#1}\fi
\expandafter\ifx\csname url\endcsname\relax
  \def\url#1{\texttt{#1}}\fi
\expandafter\ifx\csname urlprefix\endcsname\relax\def\urlprefix{URL }\fi
\providecommand{\bibinfo}[2]{#2}
\providecommand{\eprint}[2][]{\url{#2}}

\bibitem{boltzman} 
L. Boltzmann, 
Creeles J. {\bf 98}, 68 (1884).

\bibitem{stock} 
H. J. Stockmann, 
{\it Quantum Chaos: An Introduction}, (Cambridge University Press, Cambridge, U.K., 1999).

\bibitem{berry}
M. V. Berry, 
J. Phys. A: Math. Gen. {\bf 10}, 2083 (1977).

\bibitem[{\citenamefont{Deutsch}(1991)}]{deutsch91}
\bibinfo{author}{\bibfnamefont{J.~M.} \bibnamefont{Deutsch}},
  \bibinfo{journal}{Phys. Rev. A} \textbf{\bibinfo{volume}{43}},
  \bibinfo{pages}{2046} (\bibinfo{year}{1991}).

\bibitem[{\citenamefont{Srednicki}(1994)}]{srednicki94}
\bibinfo{author}{\bibfnamefont{M.}~\bibnamefont{Srednicki}},
  \bibinfo{journal}{Phys. Rev. E} \textbf{\bibinfo{volume}{50}},
  \bibinfo{pages}{888} (\bibinfo{year}{1994}).

\bibitem{cazalilla}
M. A. Cazalilla and M. Rigol,
New J. Phys. {\bf 12}, 055006 (2010);
A. Polkovnikov, K. Sengupta, A. Silva, and M. Vengalattore,
Rev. Mod. Phys. {\bf 83}, 863 (2011).

\bibitem{kinoshita06}
\bibinfo{author}{\bibfnamefont{T.}~\bibnamefont{Kinoshita}},
  \bibinfo{author}{\bibfnamefont{T.}~\bibnamefont{Wenger}}, \bibnamefont{and}
  \bibinfo{author}{\bibfnamefont{D.~S.} \bibnamefont{Weiss}},
  \bibinfo{journal}{Nature (London)} \textbf{\bibinfo{volume}{440}},
  \bibinfo{pages}{900} (\bibinfo{year}{2006}).

\bibitem[{\citenamefont{Rigol et~al.}(2007)\citenamefont{Rigol, Dunjko,
  Yurovsky, and Olshanii}}]{rigol07STAT}
\bibinfo{author}{\bibfnamefont{M.}~\bibnamefont{Rigol}},
  \bibinfo{author}{\bibfnamefont{V.}~\bibnamefont{Dunjko}},
  \bibinfo{author}{\bibfnamefont{V.}~\bibnamefont{Yurovsky}}, \bibnamefont{and}
  \bibinfo{author}{\bibfnamefont{M.}~\bibnamefont{Olshanii}},
  \bibinfo{journal}{Phys. Rev. Lett.} \textbf{\bibinfo{volume}{98}},
  \bibinfo{pages}{050405} (\bibinfo{year}{2007}); 
  M. Rigol, A. Muramatsu, and M. Olshanii, Phys. Rev. A {\bf 74}, 053616 (2006). 

\bibitem[{\citenamefont{Rigol et~al.}(2008)\citenamefont{Rigol, Dunjko, and
  Olshanii}}]{rigol08STAT}
\bibinfo{author}{\bibfnamefont{M.}~\bibnamefont{Rigol}},
  \bibinfo{author}{\bibfnamefont{V.}~\bibnamefont{Dunjko}}, \bibnamefont{and}
  \bibinfo{author}{\bibfnamefont{M.}~\bibnamefont{Olshanii}},
  \bibinfo{journal}{Nature (London)} \textbf{\bibinfo{volume}{452}},
  \bibinfo{pages}{854} (\bibinfo{year}{2008}).

\bibitem[{\citenamefont{Rigol}(2009{\natexlab{a}})}]{rigol09STAT}
\bibinfo{author}{\bibfnamefont{M.}~\bibnamefont{Rigol}},
  \bibinfo{journal}{Phys. Rev. Lett.} \textbf{\bibinfo{volume}{103}},
  \bibinfo{pages}{100403} (\bibinfo{year}{2009}{\natexlab{a}});
  \bibinfo{journal}{Phys. Rev. A} \textbf{\bibinfo{volume}{80}},
  \bibinfo{pages}{053607} (\bibinfo{year}{2009}{\natexlab{b}}).

\bibitem{santos}
  \bibinfo{author}{\bibfnamefont{L.~F.} \bibnamefont{Santos}}
  \bibnamefont{and} \bibinfo{author}{\bibfnamefont{M.}~\bibnamefont{Rigol}},
   Phys. Rev. E {\bf 81}, 036206 (2010); Phys. Rev. E {\bf 82}, 031130 (2010).

\bibitem{Neuenhahn10}
C. Neuenhahn and F. Marquardt, arXiv:1007.5306;
G. Roux, Phys. Rev. A {\bf 81}, 053604 (2010).

\bibitem{huse}
D. M. Basko, I. L. Aleiner, and B. L. Altshuler, 
Ann. Phys. {\bf 321}, 1126 (2006);
V. Oganesyan and D. A. Huse,
Phys. Rev. B {\bf 75}, 155111 (2007);
A. Pal and D. A. Huse,
Phys. Rev. B {\bf 82}, 174411 (2010); 
C. Monthus and T. Garel, Phys. Rev. B {\bf 81}, 134202 (2010).

\bibitem{fazio}
C. Gogolin, M. P. Muller, and J. Eisert,
Phys. Rev. Lett. {\bf 106}, 040401 (2011);
E. Canovi, D. Rossini, R. Fazio, G. E. Santoro, A. Silva,  
Phys. Rev. B {\bf 83}, 094431 (2011).

\bibitem{mirlin}
A. D. Mirlin, Y. V. Fyodorov, F.-M. Dittes, J. Quezada, and T. H. Seligman, 
Phys. Rev. E {\bf 54}, 3221 (1996); 
E. Cuevas, M. Ortu\~{n}o, V. Gasparian, and A. P\'erez-Garrido, 
Phys. Rev. Lett. {\bf 88}, 016401 (2001);
I. Varga, Phys. Rev. B {\bf 66}, 094201 (2002).

\bibitem{nishigaki} 
S. M. Nishigaki, 
Phys. Rev. E {\bf 59}, 2853 (1999).

\bibitem{sko} B. I. Shklovskii, B. Shapiro, B. R. Sears, P. Lambrianides, and H. B. Shore, 
Phys. Rev. B {\bf 47}, 11487 (1993); 
A. M. Garcia-Garcia and E. Cuevas, 
{\it ibid.} {\bf 75}, 174203 (2007).

\bibitem{levitov}
L. S. Levitov, Phys. Rev. Lett. 64, 547 (1990) .

\bibitem{C99}
E. Cuevas, Phys. Rev. Lett. {\bf 83}, 140 (1999),
E. Cuevas, E. Louis, and J. A. Verg\'es, {\it ibid.} {\bf 77}, 1970 (1996).

\bibitem{note1} Since we work at fixed number of fermions $p$, 
$\langle \hat{N}(k=0)\rangle=p^2/L$, so we set it to zero without any loss 
of generality.

\bibitem{dmrg}
\bibinfo{author}{\bibfnamefont{C.}~\bibnamefont{Kollath}},
  \bibinfo{author}{\bibfnamefont{A.~M.} \bibnamefont{L{\"a}uchli}},
  \bibnamefont{and} \bibinfo{author}{\bibfnamefont{E.}~\bibnamefont{Altman}},
  \bibinfo{journal}{Phys. Rev. Lett.} \textbf{\bibinfo{volume}{98}},
  \bibinfo{pages}{180601} (\bibinfo{year}{2007}); 
\bibinfo{author}{\bibfnamefont{S.~R.} \bibnamefont{Manmana}},
S. Wessel, R. M. Noack, and A. Muramatsu,
  {\it ibid.} \textbf{\bibinfo{volume}{98}},
  \bibinfo{pages}{210405} (\bibinfo{year}{2007}).

\bibitem{note2} In this regime, based on the comparison between results obtained 
for different system sizes, the saturation observed for sufficiently long times 
is expected to be a finite-size effect. 

\bibitem{cantori} 
R. S. MacKay, J. D. Meiss, and I. C. Percival,
Phys. Rev. Lett. {\bf 52}, 697 (1984);
W. Li and P. Bak, {\it ibid.} {\bf 57}, 655 (1986).

\bibitem{antwan}
A. M. Garcia-Garcia and J. Wang,
Phys. Rev. Lett. {\bf 94}, 244102 (2005); D. Wintgen and H. Marxer, 
{\it ibid.} {\bf 60}, 971 (1988); B. Hu, B. Li, J. Liu, and Y. Gu, {\it ibid.} {\bf 82}, 4224 (1999). 

\bibitem{alder}
See, for example, B. J. Alder and T. E. Wainwright,
Phys. Rev. A {\bf 1}, 18 (1970).

\bibitem{uli} S. Trotzky, Y.-A. Chen, A. Flesch, I. P. McCulloch, U.
Schollw\"ock, J. Eisert, 
and I. Bloch, Nature Phys. {\bf 8}, 325 (2012). 

\end{thebibliography}
\end{document}